\documentclass[useAMS,usenatbib]{mn2e}
\usepackage{graphicx}
\usepackage{amssymb}

\title{A magnetar powering the ordinary monster GRB 130427A?}

\author[Bernardini et al.]{M.G. Bernardini$^{1}$\thanks{E-mail: grazia.bernardini@brera.inaf.it}, S. Campana$^{1}$, G. Ghisellini$^{1}$, P. D'Avanzo$^{1}$, G. Calderone$^{1}$, \newauthor S. Covino$^{1}$, G. Cusumano$^{2}$, G. Ghirlanda$^{1}$, V. La Parola$^{2}$, A. Maselli$^{2}$, \newauthor A. Melandri$^{1}$, R. Salvaterra$^{3}$, D. Burlon$^{4,5}$, V. D'Elia$^{6}$, D. Fugazza$^{1}$, \newauthor B. Sbarufatti$^{1}$, S.D. Vergani$^{7}$, G. Tagliaferri$^{1}$\\
$^{1}$INAF - Osservatorio Astronomico di Brera, via Bianchi 46, I-23807 Merate (LC), Italy\\
$^2$ INAF -- IASF Palermo, via Ugo La Malfa 153, I-90146 Palermo, Italy\\
$^3$ INAF -- IASF Milano, via E. Bassini 15, I-20133 Milano, Italy\\
$^4$ Sydney Institute for Astronomy, School of Physics, The University of Sydney, NSW 2006, Australia\\
$^5$ ARC Centre of Excellence for All-sky Astrophysics (CAASTRO), The University of Sydney, NSW 2006, Australia\\
$^6$ INAF -- Osservatorio Astronomico di Roma, via Frascati 33, I-00040 Monteporzio Catone (RM), Italy\\
$^7$ GEPI -- Observatoire de Paris, CNRS UMR 8111, Univ. Paris-Diderot, 5 Place Jules Jannsen, F-92190 Meudon, France}
\date{\today}

\pagerange{\pageref{firstpage}--\pageref{lastpage}} \pubyear{2013}

\begin{document}

\label{firstpage}

\maketitle

\begin{abstract}
We present the analysis of the extraordinarily bright Gamma-Ray Burst (GRB) 130427A under the hypothesis that the GRB central engine is an accretion--powered magnetar. In this framework, initially proposed to explain GRBs with precursor activity, the prompt emission is produced by accretion of matter onto a newly--born magnetar, and the observed power is related to the accretion rate. The emission is eventually halted if the centrifugal forces are able to pause accretion. We show that the X-ray and optical afterglow is well explained as the forward shock emission with a jet break plus a contribution from the spin--down of the magnetar. Our modelling does not require any contribution from the reverse shock, that may still influence the afterglow light curve at radio and mm frequencies, or in the optical at early times. We derive the magnetic field ($B\sim 10^{16}$ G) and the spin period ($P\sim 20$ ms) of the magnetar and obtain an independent estimate of the minimum luminosity for accretion. This minimum luminosity results well below the prompt emission luminosity of GRB 130427A, providing a strong consistency check for the scenario where the entire prompt emission is the result of continuous accretion onto the magnetar. This is in agreement with the relatively long spin period of the magnetar. GRB 130427A was a well monitored GRB showing a very standard behavior and, thus, is a well--suited benchmark to show that an accretion--powered magnetar gives a unique view of the properties of long GRBs.
\end{abstract}


\section{Introduction}

The picture of the prompt and afterglow emission provided by the \textit{Swift} \citep{2005ApJ...621..558G} and \textit{Fermi} missions in recent years strengthened the idea that long duration Gamma-Ray Bursts (GRBs) originate from the death of massive stars \citep{2006ARA&A..44..507W}. Nevertheless, it remains unsettled whether the central engine is a rapidly accreting black hole \citep{1993ApJ...405..273W,1999ApJ...524..262M,2008MNRAS.388.1729K} or a ``millisecond magnetar'' \citep{1992Natur.357..472U,1992ApJ...392L...9D,2007ApJ...669..546U,2011MNRAS.413.2031M}, i.e. a neutron star endowed with a large magnetic field ($B\sim 10^{15-16}$ G). 

In \citet{2013arXiv1306.0013B} we proposed that GRB prompt emission originates from an accretion disk feeding a newly born magnetar, and therefore the observed power is proportional to the mass accretion rate. In this framework, close to the surface of the magnetar, the behavior of the in--falling material is dominated by the large magnetic field of the neutron star, so that matter is channelled along the field lines onto the magnetic polar caps. The magnetic field begins to dominate the motion of matter at the magnetospheric radius $r_{\rm m}$, defined by the pressure balance between the magnetic dipole of the magnetar and the in--falling material. Accretion onto the surface of the magnetar proceeds as long as the material in the disk rotates faster than the magnetosphere. In the opposite case, accretion can be substantially reduced due to centrifugal forces exerted by the super-Keplerian magnetosphere: the source is said to enter the ``propeller'' phase \citep{1975A&A....39..185I,1998A&ARv...8..279C}. This scenario has been proposed to explain the observational features of the prompt emission of GRBs with precursors, that constitute $\sim 17\%$ of entire GRB population (see also \citealt{2008ApJ...685L..19B}). Both the precursor(s) and the main emission correspond to the accretion episodes. When the system enters the propeller phase, accretion is inhibited and the GRB becomes quiescent. If we interpret the X--ray plateau in the GRB afterglows as due to continuous  energy injection from the magnetar spin--down power \citep{1998A&A...333L..87D,2001ApJ...552L..35Z,2009ApJ...702.1171C,2011A&A...526A.121D}, we can estimate the magnetic field and spin period of the magnetar and have a completely independent test for the precursor and main event nature. This model has a more general application if we consider that all GRBs with a shallow decay phase are produced by a power-accretion magnetar, independently from the presence of precursors.

In this Letter we show that an accretion--powered magnetar is a suitable central engine for the super--bright GRB 130427A, a nearby ($z=0.34$, \citealt{2013GCN..14455...1L}) GRB detected simultaneously by several high--energy satellites \citep{2013arXiv1311.5623F,2013arXiv1311.5581P,2013GCN..14487...1G}. Its extremely bright afterglow was followed up at lower energy, allowing to test standard models with great detail \citep{2013arXiv1305.2453L,2013arXiv1311.5254M,2013arXiv1307.4401P,2013MNRAS.436.3106P,2013arXiv1311.5489V}. Overall, GRB 130427A shares the properties of cosmological GRBs \citep{2013arXiv1311.5254M}, and it is the first case of a powerful GRB associated with a Type Ic Supernova (SN) (\citealt{2013arXiv1305.6832X}, Melandri et al. 2013, in preparation). Indeed, all the other cases were found to be associated to GRBs with peak luminosity $L< 10^{51}$ ergs s$^{-1}$(see e.g. GRB 980425/SN 1998bw, \citealt{1999A&AS..138..463P}, or GRB 060218/SN 2006aj, \citealt{2006Natur.442.1008C,2006Natur.442.1014S}). In \citet{2013arXiv1306.0013B} we envisaged that both low--luminosity and normal GRBs can originate from the collapse of a massive star leaving a magnetar. In the first case, the mass inflow rate is not high enough to shut off the spin--down power and we have a GRB powered by the wind of the magnetar rather than by accretion as for normal GRBs. GRB 130427A, bridging the gap between nearby, low--luminosity and powerful, cosmological GRBs is a benchmark to test the scenario proposed in \citet{2013arXiv1306.0013B}.

\begin{figure*}
\centering
\includegraphics[width=0.85\hsize,clip]{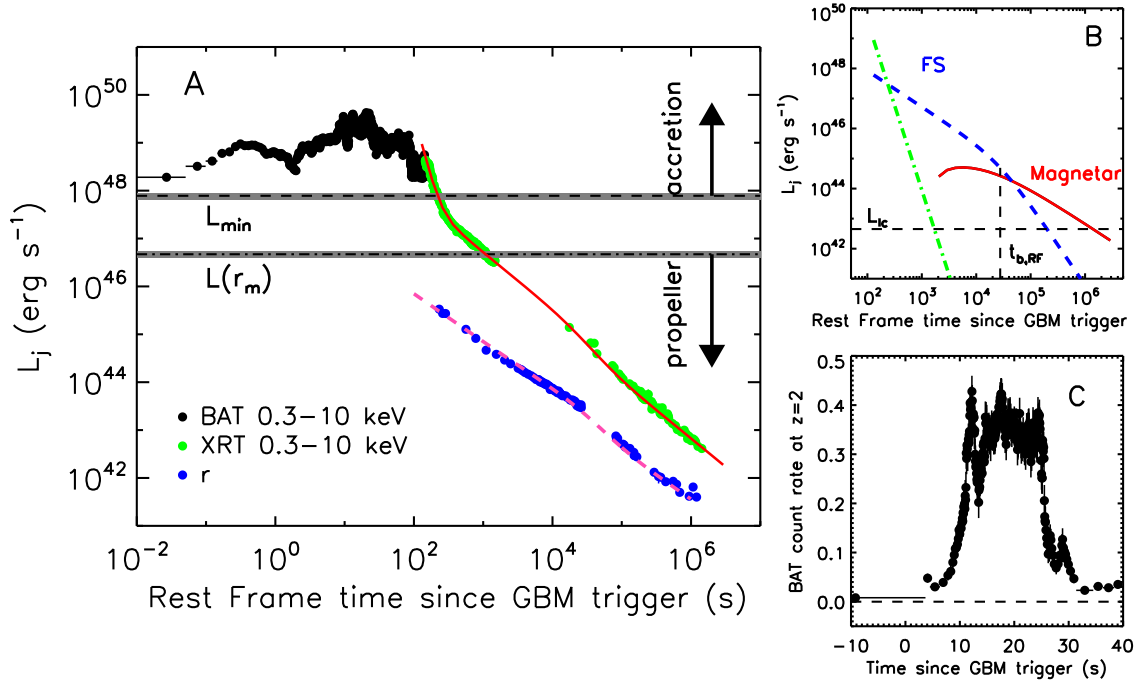}
\caption{X--ray and optical light curves of GRB 130427A (panel A). Luminosity lines $k$--corrected in the $0.3-10$ keV observed energy band are compared to the prompt emission (black points), that is above the estimate for $L_{\rm min}$. The gray areas mark the $1 \sigma$ region around $L(r_{\rm m})$ and $L_{\rm min}$. The red solid line marks the best fit to the afterglow in the $0.3-10$ keV energy band (green points), while the pink dashed line is the model luminosity $k$-corrected in the optical band (\textit{r'} filter) compared to observations (blue points). The host galaxy contribution has been subtracted. The three different components in the X--ray afterglow emission (panel B): the initial steep decay (green dash--dotted line), the forward shock emission (FS, blue dashed line) and the contribution from the wind of the magnetar (Magnetar, red solid line). The end of the accretion process corresponds to the moment when the accretion power (the green dash--dotted line) falls below the luminosity at the light cylinder $L_{\rm lc}$. From this time on, the magnetar start to contribute to the afterglow emission with its spin--down power (red solid line). After $t_{\rm b,RF}=27.6$ ks the slope of the FS changes due to the jet break. Mask--weighted BAT count rate light curve of the first peak of GRB 130427A as it would appear at $z=2$ (panel C).}
\label{plot}
\end{figure*}

\section{GRB 130427A discovery and observations}\label{sect_obs}

GRB 130427A was discovered by the \textit{Fermi}/Gamma--ray Burst Monitor \citep[GBM,][]{2009ApJ...702..791M} at $07:47:06.42$ UT on April 27 2013 \citep{2013arXiv1311.5581P}. Hereafter, this time will be our reference time $T_\circ$. It was also detected by Konus--Wind \citep{2013GCN..14487...1G} and by the \textit{Swift}/Burst Alert Telescope \citep[BAT,][]{2005SSRv..120..143B}, $50.6$ s after the \textit{Fermi} trigger \citep{2013arXiv1311.5254M}. With its extremely large peak flux \citep[$f_{pk,1s}=331$ ph/(cm$^2$s) in the BAT $15-150$ keV energy band;][]{2013arXiv1311.5254M}, GRB 130427A is the brightest GRB observed by \textit{Swift} and \textit{Fermi}. High--energy gamma--ray emission was detected by the \textit{Fermi}/Large Area Telescope \citep[LAT,][]{2009ApJ...697.1071A} up to $94$ GeV \citep{2013arXiv1311.5623F}.

The \textit{Swift}/X--ray Telescope (XRT, \citealt{2005SSRv..120..165B}) began observing the field $195$ s after the GBM trigger, leading to the detection and localization of the X-ray afterglow \citep{2013arXiv1311.5254M}. The \textit{Swift}/UV/Optical Telescope \citep[UVOT,][]{2005SSRv..120...95R} began observing the field $181$ s after the trigger, leading to the detection of a bright UV/optical afterglow \citep{2013arXiv1311.5254M} which was detected by several ground-based observatories in the optical, millimeter, and radio bands. In particular, a bright optical counterpart was promptly detected by the Raptor robotic telescope \citep{2013arXiv1311.5489V}.

The redshift was found to be $z = 0.34$ from optical spectroscopy of the afterglow \citep{2013GCN..14455...1L}. Its relative proximity made this GRB the ideal candidate to study the association with SNe. Indeed the broad line Type Ic SN 2013cq was associated with this burst \citep{2013arXiv1305.6832X} and followed extensively for a very long period after the explosion (Melandri et al. 2013, in preparation). Thus, this is the first case of a powerful GRB associated with a Type Ic SN.

Its extremely bright and well--monitored afterglow allows to test standard models with great detail. \citet{2013arXiv1311.5254M} found that from $26.6$ ks the X--ray and optical light curves show the same temporal behavior, well fitted by a forward shock model in a homogeneous circumburst medium. Before this time, an extra--component in the X--ray energy band is required, probably related to the MeV prompt emission \citep{2013arXiv1311.5254M}.  The richness of the afterglow dataset allowed \citet{2013arXiv1311.5254M} to identify an achromatic break at $\sim 37$ ks, suggestive of a jet break. 

Alternative models have been proposed to interpret the multiwavelength afterglow emission (from radio to GeV). \citet{2013arXiv1305.2453L} and \citet{2013arXiv1307.4401P} proposed a (Newtonian) reverse plus forward shock model: the reverse shock is responsible for the radio and mm emission and, partially, for the optical emission at early times, while the forward shock emission is responsible for the optical and X-ray emission. \citet{2013arXiv1311.5489V} advocated the reverse shock emission to explain the initial optical flash observed by Raptor and the initial optical light curve  \citep[see also][]{2013MNRAS.436.3106P}. All these proposals share the common idea of a wind--like circumburst medium, and that the entire X--ray emission is from the forward shock only.

\section{GRB 130427A in the context of the accretion--powered magnetar model}\label{sect_model}

Following the model described in \citet{2013arXiv1306.0013B}, we interpret the GRB 130427A prompt emission as originated from a newly born magnetar accreting material from an accretion disk onto its surface. This proceeds as long as the material in the disk rotates faster than the magnetosphere, and the emission should lie above the characteristic luminosity corresponding to the onset of the propeller phase $L_{\rm min}$. The accretion process ends when the mass inflow rate decreases enough for the magnetospheric radius to reach the light cylinder (i.e. the radius at which the field lines co--rotate with the neutron star at the speed of light). Beyond this radius, i.e. when the accretion power falls below the luminosity at the light cylinder radius $L_{\rm lc}$, the field becomes radiative and expels much of the in--falling matter. For larger distances, the GRB afterglow can also be influenced by the magnetar, being re--energized by its spin--down power \citep{2011A&A...526A.121D}.

If during the accretion phase the magnetar accretes enough matter (as we proposed has occurred in GRB 061007, see \citealt{2013arXiv1306.0013B} for a discussion), then the compact object collapses to a black hole \citep{2011ApJ...736..108P}. The large isotropic energy of GRB 130427A should be rescaled by the beaming factor $f_b=(1-\cos \theta_j)$, that can be inferred from the optical and X-rays observations. The jet break at $t_{\rm b}\sim 37$ ks ($t_{\rm b,RF}=27.6$ ks rest frame) corresponds to a collimation angle $\theta_j\sim 3.4^\circ$ \citep{2013arXiv1311.5254M}. Thus, the total bolometric kinetic energy is $E_j =(f_b/\eta)\, E_{\gamma,{\rm iso}}\sim 2\times 10^{52}$ erg (assuming a radiative efficiency $\eta=0.1$\footnote{We adopt standard values of the cosmological parameters: $H_\circ=70$ km s$^{-1}$ Mpc$^{-1}$, $\Omega_M=0.27$, and $\Omega_{\Lambda}=0.73$.}). This energy corresponds to an accreted mass $M_{\rm acc,j}\sim E_j/c^2 \sim 0.02$ M$_\odot$. The small amount of accreted mass suggests that the magnetar likely did not collapse to a black hole at the end of the prompt emission. 

Consistently with the analysis of the X--ray and optical data reported in \citet{2013arXiv1311.5254M}, we considered the X--ray emission as the afterglow emission produced by the forward shock with a jet break at $t_{\rm b,RF}=27.6$ ks. The X-ray emission is not a simple power-law but shows a curvature that cannot be fully captured by a simple forward shock emission (see \citealt{2013arXiv1311.5254M} their fig.~S7 and \citealt{2013arXiv1307.4401P} their fig.~11 where the extrapolation backwards of their forward shock model underestimates and overestimates, respectively, the X-ray emission). We therefore propose that the magnetar contributes to the afterglow emission with its spin--down power. According to the scenario outlined in \citet{2011A&A...526A.121D}, the afterglow emission is the sum of the forward shock emission (a power law) plus the contribution of the wind of the magnetar, and this should account for the X--ray light curve from $\sim 200$ s onward \citep[see][their Eq.~8]{2011A&A...526A.121D}. 

Since the spin--down power depends on the magnetic field and spin period, we can obtain a direct estimate of these parameters from the best fit to the X--ray data. We introduced two modifications to the function derived by \citet{2011A&A...526A.121D}: we adopted for the forward shock emission a broken power law to account for the jet break occurring at $t_{\rm b,RF}=27.6$ ks. We added a further power--law component ($A\,(t/$s$)^{-a}$) to model the initial steep decay of the X--ray light curve, that is ascribed to the end of the accretion and, consequently, of the prompt emission. The three components of the X--ray emission are illustrated in fig.~\ref{plot}. 

The best fit to the X--ray data is portrayed in fig.~\ref{plot}. The magnetic field $B=(1.15\pm 0.09)\times 10^{16}$ G and the spin period $P=(24.2\pm 0.2)$ ms\footnote{Errors are given at $1\, \sigma$ confidence level.} resulting from the fit are comparable to the distributions of these two parameters for GRBs without precursors (see \citealt{2013arXiv1306.0013B}, therein fig.~4). The other free parameters are the initial forward shock energy $E_\circ=(2.07\pm 0.01)\times 10^{50}$ erg, the normalization of the initial steep power law $A=(2.24\pm 0.72)\times 10^{61}$ erg s$^{-1}$ and its power--law index $a=(5.65\pm 0.06)$. The initial time when the spin--down power of the magnetar starts to contribute to the afterglow emission is fixed to $T_{\circ,RF}=1500$ s, that is the time when the accretion power falls below $L_{\rm lc}$ and the accretion process ends (see fig.~\ref{plot}). We also fixed the parameter $k'$ to be $k'=0.2$ in order to get the best agreement with the X-ray late time emission (the light curve asymptotical behaviour is $\propto t^{-(1+k')}$). The parameter $k'$ is bracketed between 0 and 1 and depends on the fraction of energy transferred to the electrons and on the evolution of the bulk Lorentz factor, that in turn depends on the ambient medium (ISM or wind, see \citealt{2011A&A...526A.121D} for further details). Since, as a general trend, larger values of $k'$ are expected for wind-like medium (see \citealt{2011A&A...526A.121D}), the physical implications of such a low value of $k'$ are that an ISM medium is favored. This value is at variance with the interpretations of the afterglow based on the reverse shock emission \citep{2013arXiv1305.2453L,2013arXiv1307.4401P,2013MNRAS.436.3106P,2013arXiv1311.5489V}, while it is in agreement with the scenario proposed by \citet{2013arXiv1311.5254M}. The post jet break slope variation of the forward shock component is fixed to $\Delta=1.5$, consistent with the expectations in the standard afterglow model \citep{1999ApJ...519L..17S}. However, the observed afterglow light curve after the jet break is shallower than expected because of the injection of energy from the wind of the magnetar. An injection of energy was also envisaged by \citet{2013arXiv1311.5254M} as a possible explanation for the shallower--than--expected post--break slope\footnote{Alternatively, this unconventional post--break shallow decay could be ascribed to the time evolution of the microphysical parameters \citep{2013arXiv1311.5254M}.}. 
 
The X--ray and optical light curves have a similar behavior from $26.6$ ks onward, while at earlier times the optical light curve is shallower. We considered the best fit to X--ray data without the steep power--law component used to model the initial decay (which is likely related to the tail of the prompt emission and not to the afterglow emission). The remaining two components (the forward shock emission and the spin--down power of the magnetar) have been $k$-corrected in the \textit{r'} filter using the spectral analysis of the X--ray data reported in \citet[][Supplementary Table 4]{2013arXiv1311.5254M}. We overlaid our model light curve without any additional fit on the observations by the Faulkes Telescope North and by the Liverpool Telescope finding a good match (see fig.~\ref{plot}), as expected since there are no spectral breaks between the X--ray and optical bands \citep{2013arXiv1311.5254M}. From our analysis of the X-ray emission it turns out that also the optical emission after $t\sim 400$ s can be accounted for by the forward shock with the contribution of the magnetar and, thus, does not require any additional contribution from the reverse shock.

With the direct and independent estimate of the magnetic field and spin period of the magnetar from the fit to the X--ray emission, we calculated the characteristic bolometric luminosity corresponding to the onset of the propeller phase, $L_{\rm min}=3.1\times 10^{49}$ erg s$^{-1}$, the luminosity during the propeller phase resulting from the gravitational energy release of the in--falling matter up to $r_{\rm m}$, $L(r_{\rm m})=1.8\times 10^{48}$ erg s$^{-1}$, and the luminosity at the light cylinder radius, $L_{\rm lc}=1.81\times 10^{44}$ erg s$^{-1}$. In fig.~\ref{plot} we compare these luminosities scaled in the XRT $0.3-10$ keV energy band with the prompt emission rescaled in the same energy band. The observed luminosity is consistently above the minimum luminosity expected for accretion, thus confirming that the entire prompt emission is the result of continuous accretion onto the magnetar. This is supported observationally by the fact that the spectral index evolves during the burst without discontinuity \citep[][therein fig.~1]{2013arXiv1311.5254M}. The large spin period leads to weaker centrifugal forces and the magnetar never enters the propeller phase during the prompt emission, and in fact we do not observe quiescent times. 

Since this is a nearby GRB, one may ask if a burst with similar characteristics that would have exploded at cosmological distances would have been misclassified as a GRB with a precursor. In this case, the quiescent time would have been related to the undetectability of the emission, and not to the properties of the magnetar and of the accretion flow. We checked this rescaling the BAT light curve as it would appear at redshift $z\sim2$, that is a typical redshift for the bright GRBs observed by \textit{Swift} \citep[the BAT6 sample,][]{2012ApJ...749...68S}. Adopting the spectral analysis performed in \citet[][therein Supplementary Table 4]{2013arXiv1311.5254M}, we normalized each spectrum in order to obtain the expected flux at $z=2$ and we shifted the energy band in order to account for the cosmological redshift of the frequencies. We used these scaled spectra to simulate with XSPEC (ver. 12.6.1) how BAT would have detected it, and we used the \texttt{batphasimerr} task to create the correct uncertainties. The mask--weighted BAT light curve at $z=2$ is portrayed in fig.~\ref{plot}. For consistency with the procedure to identify the precursors in Bernardini et al. (2013), we rebinned the light curve at $S/N=5$. The emission between the first and the main peak is still detectable. Thus, even at $z=2$, this burst would not have shown precursor activity. This supports the consistency of the classification of GRBs with and without precursors in \citet{2013arXiv1306.0013B}. 

The possibility of a magnetar--powered SN via a propeller mechanism has been explored by \citet{2011ApJ...736..108P}. They concluded that for small, hydrogen--poor envelopes a broad--lined Type Ic SN may appear, similar to SN 2013cq that has been detected associated with this burst \citep{2013arXiv1305.6832X}. Therefore, the proposed model can easily accommodate both the GRB emission and the associated SN.

\section{Conclusions}

In this Letter we analyzed the prompt and afterglow emission of GRB 130427A, a nearby, powerful GRB recently discovered and observed with unprecedented details. An energy injection from the magnetar is needed to interpret the entire X-ray emission, that is not a simple power-law but shows a curvature that cannot be fully captured by the forward shock emission only. From our analysis of the X-ray emission it turns out that also the early optical emission can be accounted for by the forward shock with the contribution of the magnetar.

From the X--ray light curve fit we can derive the magnetic field and the spin period of the magnetar and, thus, an independent estimate of the minimum luminosity for accretion $L_{\rm min}$. When compared to the prompt emission luminosity of GRB 130427A, this results well above $L_{\rm min}$, confirming that the entire prompt emission is the result of continuous accretion onto the magnetar. This implies that the absence of quiescent times is intrinsic (i.e. the source does not enter the propeller phase) and not due to the capability to detect the emission among different peaks because of its relative proximity. This model requires that the late X-ray and optical light curves curvature is interpreted as the combination of energy injection and jet break, while it does not require any contribution from the reverse shock. However, reverse shock emission may still influence the afterglow light curve at radio and mm frequencies \citep{2013arXiv1305.2453L,2013arXiv1307.4401P}, or during the very early optical flash observed by Raptor \citep{2013arXiv1311.5489V}. GRB 130427A revealed itself as an important benchmark to show that an accretion--powered magnetar is able to explain the main properties of long GRBs in general, thus extending the validity of this model initially proposed for GRBs with precursors \citep{2013arXiv1306.0013B}.

\section*{}

The authors acknowledge support from PRIN--MIUR 2009ERC3HT and ASI--INAF I/088/06/0 grants. DB acknowledges the support of the Australian Research Council through grantÊDP110102034.

\label{lastpage}

\end{document}